\documentclass[prd,aps,preprintnumbers,nofootinbib,twocolumn]{revtex4}
\usepackage{epsfig}
\usepackage{amsmath}
\usepackage{hyperref}

\def\t{{\bar t}}
\def\Mtt{m_{t\bar{t}}}

\def\as{\alpha_S}
\def\dphi{\Delta\phi_{\ell\ell}}
\def\deta{|\Delta\eta_{\ell\ell}|}
\def\GeV{{\rm GeV}}

\begin{document}

\preprint{Cavendish-HEP-19/02, TTK-19-02, TTP19-003}

\renewcommand{\thefigure}{\arabic{figure}}

\title{Higher order corrections to spin correlations in top quark pair production at the LHC}

\author{Arnd Behring}
\affiliation{{\small Institut f\"ur Theoretische Teilchenphysik und Kosmologie, RWTH Aachen University, D-52056 Aachen, Germany}}
\affiliation{{\small Institute for Theoretical Particle Physics, KIT, Karlsruhe, Germany}}
\author{Micha\l{}  Czakon}
\affiliation{{\small Institut f\"ur Theoretische Teilchenphysik und Kosmologie, RWTH Aachen University, D-52056 Aachen, Germany}}
\author{Alexander Mitov}
\author{Andrew S.~Papanastasiou}
\author{Rene Poncelet}
\affiliation{{\small Cavendish Laboratory, University of Cambridge, Cambridge CB3 0HE, UK}}

\date{\today}

\begin{abstract}
We calculate, for the first time, the next-to-next-to leading order (NNLO) QCD corrections to spin correlations in top quark pair production at the LHC. The NNLO corrections play an important role in the description of the corresponding differential distributions. We observe that the Standard Model calculation describes the available $\dphi$ data in the fiducial region but does not agree with the $\dphi$ measurement extrapolated to full phase space. Most likely this discrepancy is due to the difference in precision between existing event generators and NNLO calculations for dilepton top-pair final states. 
\end{abstract}
\maketitle

\section{Introduction}

Within the Standard Model (SM) of particle physics individual top quarks produced in proton--(anti)proton collisions are not polarized. The spins of two {\it pair-produced} top quarks are, however, correlated to each other. It is possible \cite{Barger:1988jj} to study directly such {\it spin correlations} between top quarks since, due to the very rapid decay of the top quark, its spin is passed to its decay products almost free of non-perturbative effects \cite{Bigi:1986jk}. This implies that top quark spin correlations are calculable.

The study of top quark spin correlations has a long history. Spin correlations have been long recognized as a powerful tool for probing the nature of the quark sector in the Standard Model (SM) \cite{Bernreuther:1995cx,Mahlon:1995zn,Bernreuther:2000yn,Bernreuther:2001bx,Bernreuther:2001rq,Mahlon:2010gw,Bernreuther:2010ny,Melnikov:2011ai,Bernreuther:2015yna,Bernreuther:2017yhg,Aguilar-Saavedra:2018ggp,Richardson:2018pvo} as well as Higgs and/or beyond the SM (BSM) physics \cite{Hill:1993hs,Han:2012fw,Baumgart:2012ay,Han:2013lna,Biswas:2014hwa}. Indeed, a generic BSM contribution to top production will alter the top-pair production spin density matrix. An important example is the case of a light spin zero top quark supersymmetric partner, the stop, decaying to top quarks \cite{Han:2012fw,Han:2013lna}. Seeking deviations between SM predictions and LHC measurements of top quark spin correlations represents a powerful model-independent search strategy for possible BSM physics coupled to the top quark sector.

Very recently, the ATLAS collaboration has published \cite{ATLAS:2018rgl} a very precise measurement of spin correlations in top quark pair production at the LHC (earlier LHC and Tevatron measurements include \cite{Chatrchyan:2013wua,Aad:2014pwa,Abazov:2015psg,Khachatryan:2016xws,Aaboud:2016bit}). A deviation of about $3.2\sigma$ with respect to the SM has been observed. This is by far the biggest deviation from the SM observed in the top quark sector at the LHC to date. Given the potential significance of such a discrepancy, in this work we calculate for the first time the complete set of NNLO QCD corrections to top quark pair production and decay. Our calculation uses the narrow width approximation. It allows us to qualitatively increase the level of precision of SM predictions for realistic top quark final states thus making the comparison with the ATLAS data \cite{ATLAS:2018rgl} much more predictive.

Generally, top quark spin correlations can be assessed following two strategies. The first strategy, which we call direct, reconstructs the top-pair spin density matrix and is based on kinematic distributions computed in specially designed frames of reference; see refs.~\cite{Mahlon:2010gw,Bernreuther:2010ny} for details. 

The second strategy, which we call indirect, utilizes differential distributions defined in the laboratory frame. These distributions are best suited for experimental study but they tend to be only partly sensitive to spin correlations. In order to maximize the extracted information about spin correlations the use of a likelihood function was advocated in ref.~\cite{Melnikov:2011ai}. Clearly, a prerequisite for extracting spin correlations from laboratory frame distributions is good control over theory predictions. 

In this work we use the {\it indirect} approach to spin correlations and study the following differential distributions: the angular difference $\dphi$ between the two leptons in the transverse plane and the rapidity difference $\deta$ between the two leptons. Both observables are sensitive to spin correlations and can be measured with high precision since the top quarks need not be reconstructed. 

Our main goal in this work is to establish if higher order corrections can account for the $3.2\sigma$ discrepancy in the $\dphi$ distribution reported in ref.~\cite{ATLAS:2018rgl}. Our finding is affirmative. In hindsight, this should not come as a complete surprise given the important role higher order QCD corrections play in the $t\t$ forward-backward asymmetry \cite{Czakon:2014xsa} and in taming the so-called top $p_T$ discrepancy \cite{Czakon:2015owf}. We caution, however, that the interpretation of higher order corrections is not completely straightforward since it uncovers possible subtleties in the modeling of realistic top quark final states at hadron colliders. We explain all this in detail in sec.~\ref{sec:results}.

\section{Details about the calculation}\label{sec:calculation}

The calculations performed in this work are at next-to-next-to-leading order (NNLO) in QCD. This means that NNLO QCD corrections to both top pair production and top quark decay are included. This is the first time top quark pair production and decay has been consistently computed in NNLO QCD. 

Similarly to ref.~\cite{Gao:2017goi} where top pair production was included in approximate NNLO, the present calculation is performed within the narrow width approximation for both the top quark and the $W$ boson. This approximation is known to work well \cite{Denner:2012yc} for distributions that are away from kinematic boundaries which is the case considered in this work. Below we also compare our calculation with a more recent NLO study \cite{Denner:2016jyo}. We recall that this approximation has been used in the existing NNLO QCD calculations for single top production including top decay \cite{Berger:2016oht,Berger:2017zof}.

We consistently truncate the {\it production} $\times$ {\it decay} differential cross-section through NNLO in QCD:
\begin{eqnarray}
d\sigma^{\rm LO} &=& d\sigma^{\rm LO \times LO}\,,\nonumber\\
d\sigma^{\rm NLO} &=& d\sigma^{\rm NLO \times LO} + d\sigma^{\rm LO \times NLO} -{2\Gamma_t^{(1)}\over \Gamma_t^{(0)}} d\sigma^{\rm LO}\,, \label{eq:expansion}\\
d\sigma^{\rm NNLO} &=& d\sigma^{\rm NNLO \times LO} + d\sigma^{\rm LO \times NNLO} + d\sigma^{\rm NLO \times NLO}\nonumber\\
&-&{2\Gamma_t^{(1)}\over \Gamma_t^{(0)}} d\sigma^{\rm NLO} -{\left(\Gamma_t^{(1)}\right)^2 +2\Gamma_t^{(0)}\Gamma_t^{(2)} \over \left(\Gamma_t^{(0)}\right)^2}d\sigma^{\rm LO}\,.\nonumber
\end{eqnarray}
As the above equations imply, the top quark decay width has also been expanded in powers of $\as$ (as in eq.~(\ref{eq:numbers})). The contribution containing NLO corrections to the two decays is included in $d\sigma^{\rm LO \times NNLO}$.

The $d\sigma^{\rm NLO}$ correction has been known for some time \cite{Bernreuther:2004jv,Melnikov:2009dn,Campbell:2012uf}. These results were extended in ref.~\cite{Broggio:2014yca} to include approximate NNLO results in production while ref.~\cite{Gao:2017goi} combined the approximate NNLO correction in production with the complete NNLO correction in decay.

Our calculation uses the {\tt STRIPPER} framework \cite{Czakon:2010td,Czakon:2011ve,Czakon:2014oma} for NNLO calculations in QCD. The only exception is the calculation of the $d\sigma^{\rm NLO \times NLO}$ contribution where, purely for convenience, the {\it decay} correction is computed with the help of Catani-Seymour dipoles \cite{Catani:1996jh,Catani:2002hc} as implemented in ref.~\cite{Campbell:2004ch}.

We modify the existing calculations of differential top-pair production \cite{Czakon:2015owf,Czakon:2016ckf,Czakon:2016dgf} in such a way that the information about the helicities of the top quarks is retained. For the double real correction at NNLO this requires the use of tree-level helicity amplitudes. The real-virtual corrections require the calculation of the one-loop five-point helicity amplitudes for the processes $q\bar q\to t\t g$, $q g\to t\t q$ and $gg\to t\t g$. 
%
%
To that end we have used a private version \cite{upcomingOL2} of the {\tt OpenLoops2} code \cite{Cascioli:2011va} which employs the stability and speed improvements of \cite{Buccioni:2017yxi}.
We have checked that the result agrees with a modified version of a private code by S.~Dittmaier used previously in the calculation of spin-averaged top production. The two-loop amplitudes $q\bar q\to t\t$ and $gg\to t\t$ have been computed in ref.~\cite{Chen:2017jvi} using spin projections and the methods used for the derivation of the spin-averaged amplitudes \cite{Baernreuther:2013caa}. 
  
We have computed independently the NNLO QCD correction to top decay. The two-loop helicity amplitude $t\to bW(\to \ell\nu)$ is known analytically \cite{Bonciani:2008wf,Asatrian:2008uk,Beneke:2008ei}. The one-loop helicity amplitude for $t\to bgW(\to \ell\nu)$ has been computed in analytical form in ref.~\cite{Lim:2018qiw} and has been checked numerically against the {\tt OpenLoops} and {\tt Gosam} \cite{Cullen:2014yla} libraries. Alternatively, we have implemented the results of ref.~\cite{Campbell:2004ch} and find agreement between the two. We have checked that the assembled fully differential spin-averaged top quark decay width agrees within numerical uncertainties with the program {\tt NNTopDec} \cite{Gao:2012ja} as well as with ref.~\cite{Brucherseifer:2013iv} (where such comparison was possible). 

For all terms in eq.~(\ref{eq:expansion}) but $d\sigma^{\rm NNLO \times LO}$ we have checked that they agree with ref.~\cite{Gao:2017goi}.

In the calculation we work in the $G_\mu$-scheme and use the following set of numerical inputs
\begin{eqnarray}
m_t &=& 172.5\,\GeV\,,\label{eq:numbers}\\
m_W &=& 80.385\,\GeV\,,\nonumber\\
m_Z &=& 91.1876\,\GeV\,,\nonumber\\
\Gamma_W &=& 2.0928\,\GeV\,,\nonumber\\
\Gamma_t &=& \left(1.48063 -1.18\,\as  -2.65\,\as^2\right) \,\GeV\,,\nonumber\\
G_F &=& 1.166379\times 10^{-5}\,\GeV^{-2}\,.\nonumber
\end{eqnarray}
The top quark width is specified as an $\as$-expansion through NNLO in QCD \cite{Blokland:2004ye,Gao:2012ja,Brucherseifer:2013iv} as needed in eq.~(\ref{eq:expansion}). It is treated as a fixed parameter throughout this work and its value in eq.~(\ref{eq:numbers}) corresponds to a fixed scale $\mu=m_t$.

In this work we take the $b$-quark to be massless and renormalize with $n_F=5$ active flavors. The top quark is renormalized on-shell, i.e. we use the top quark pole mass. Its value $m_t=172.5\,\GeV$ is lower than the world average. It is chosen such that it agrees with the value used by the ATLAS collaboration. This way our predictions can be directly compared with ref.~\cite{ATLAS:2018rgl}.

We use the NNPDF3.1 \cite{Ball:2017nwa} family of parton distributions but have also checked the CT14 set \cite{Dulat:2015mca}. The default renormalization and factorization scales are chosen dynamically as proposed in ref.~\cite{Czakon:2016dgf}:
\begin{equation}
\mu_{F,R} = {H_T\over 4}~, ~~H_T=\sqrt{m_t^2+p_{T,t}^2}+\sqrt{m_t^2+p_{T,\t}^2}\,.
\label{eq:scales}
\end{equation}
In the evaluation of the scales eq.~(\ref{eq:scales}) we have used the true top momenta, not the reconstructed ones. 

Although our setup allows us to output {\tt fastNLO} tables \cite{Kluge:2006xs,Britzger:2012bs} in this first NNLO calculation with top decay we have chosen for simplicity not to do so. We plan to use this capability in future calculations similarly to our results \cite{Czakon:2017dip} of stable top quark production. All results derived in this work, together with some extra plots, are available for download from \cite{web-based-results}.

\section{Results}\label{sec:results}

In this work we calculate two differential distributions, namely, the two leptons' angular difference in the transverse plane $\dphi$ and their rapidity difference $\deta$. 

We have two selection criteria for each distribution. The first one, called {\it inclusive}, does not assume any selection cuts. The second one, called {\it fiducial}, is based on the ATLAS selection cuts \cite{ATLAS:2018rgl}: an electron and a muon of opposite electric charge with $p_T > 27 (25)\,\GeV$ for the harder (softer) lepton and $|\eta|<2.5$. In addition, we require at least two jets (at least one of which is a $b$-flavored jet) with $p_T>25\,\GeV$ and $|\eta|<2.5$. All jets are defined with the anti-$k_T$ algorithm \cite{Cacciari:2008gp} with $R=0.4$.

\begin{figure}[t]
  \centering
     \hspace{-1mm} 
     \includegraphics[width=0.48\textwidth]{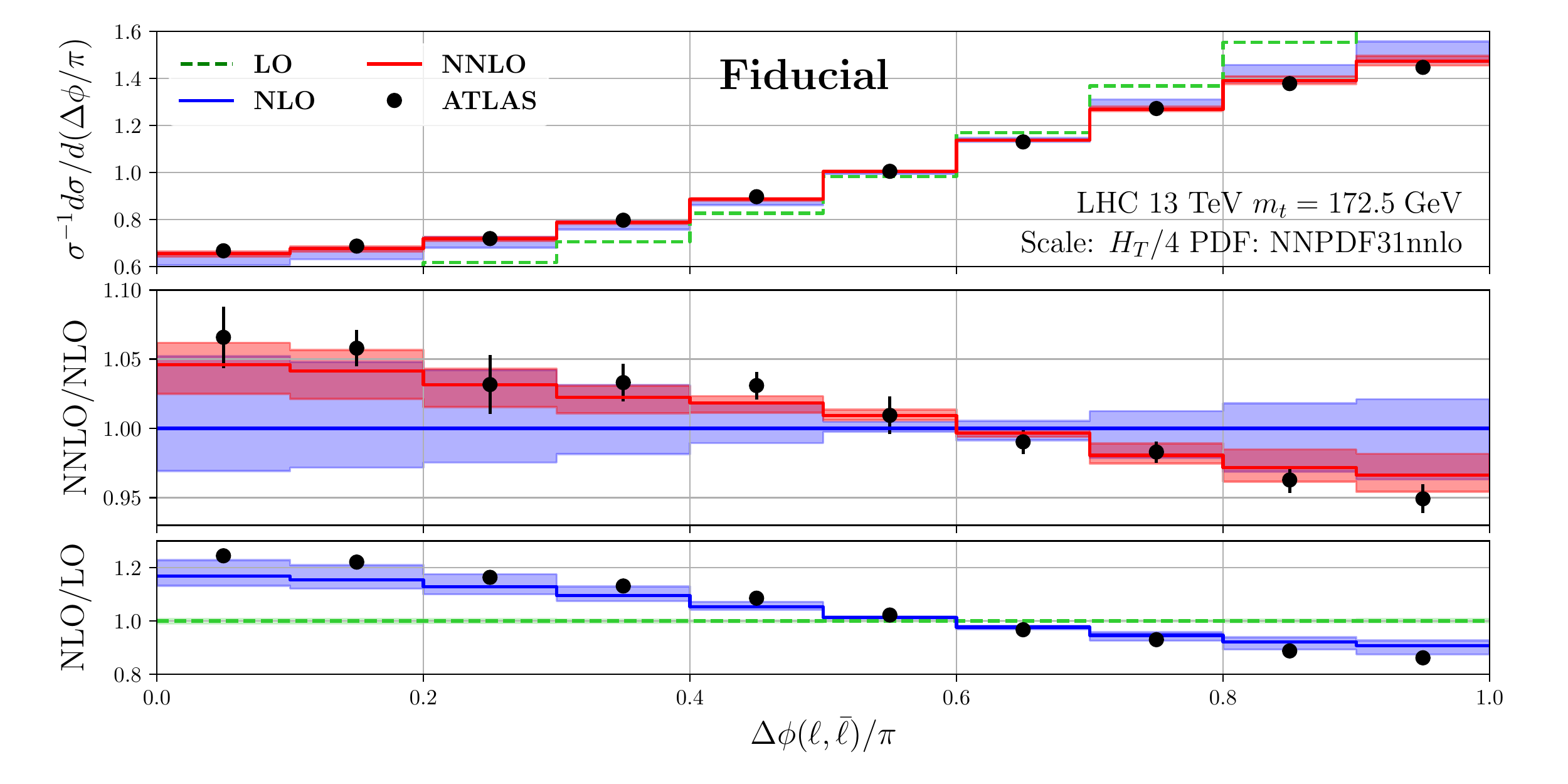}
     \includegraphics[width=0.48\textwidth]{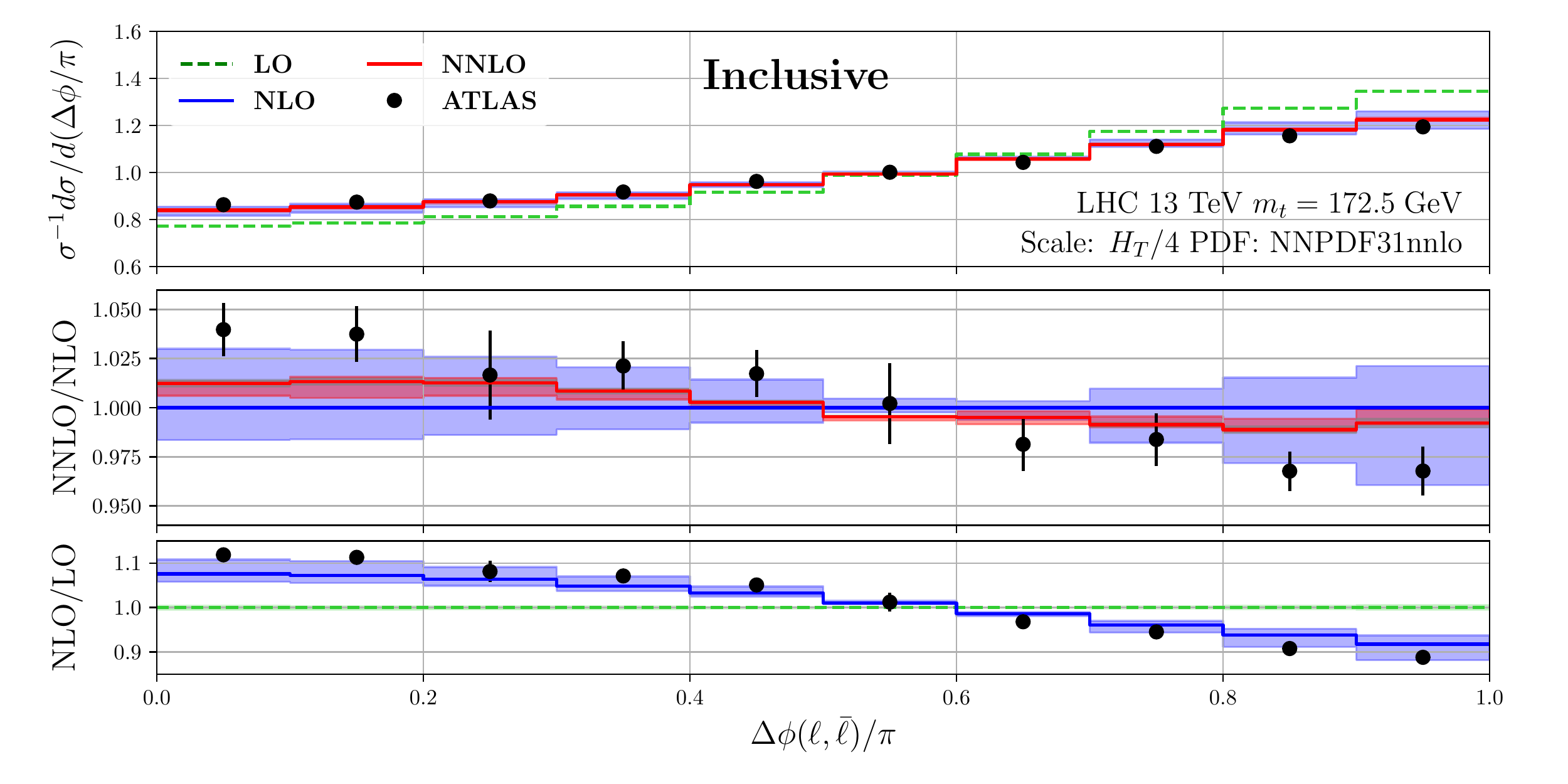}  
   \caption{NNLO QCD predictions for the fiducial (top) and inclusive selections (bottom) of the normalized $\dphi$ distribution versus ATLAS data \cite{ATLAS:2018rgl}. Uncertainty bands are from 7-point scale variation.}
   \label{fig:dphi}
\end{figure}

\begin{figure}[t]
  \centering
     \hspace{-1mm} 
     \includegraphics[width=0.48\textwidth]{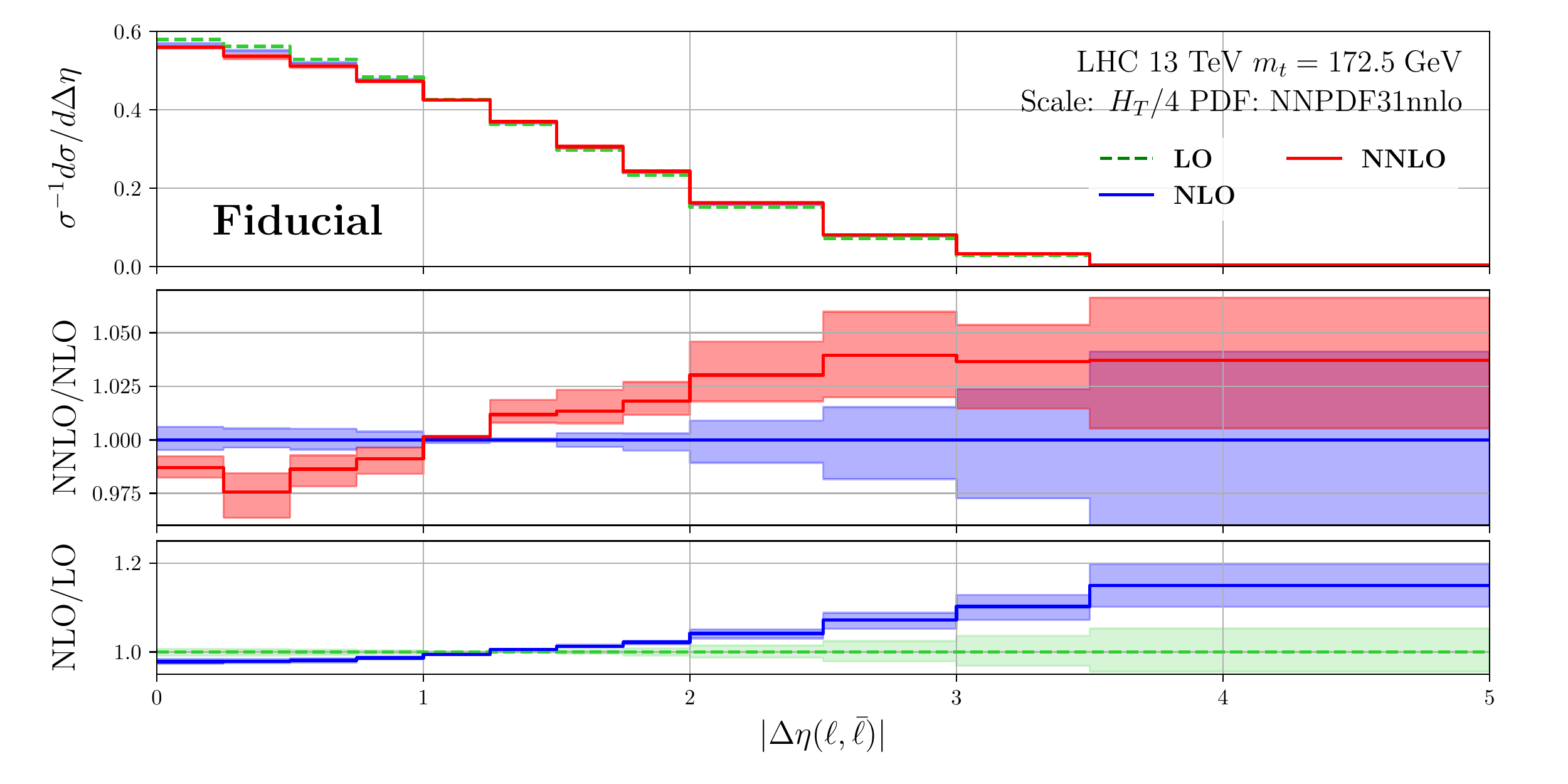}
     \includegraphics[width=0.48\textwidth]{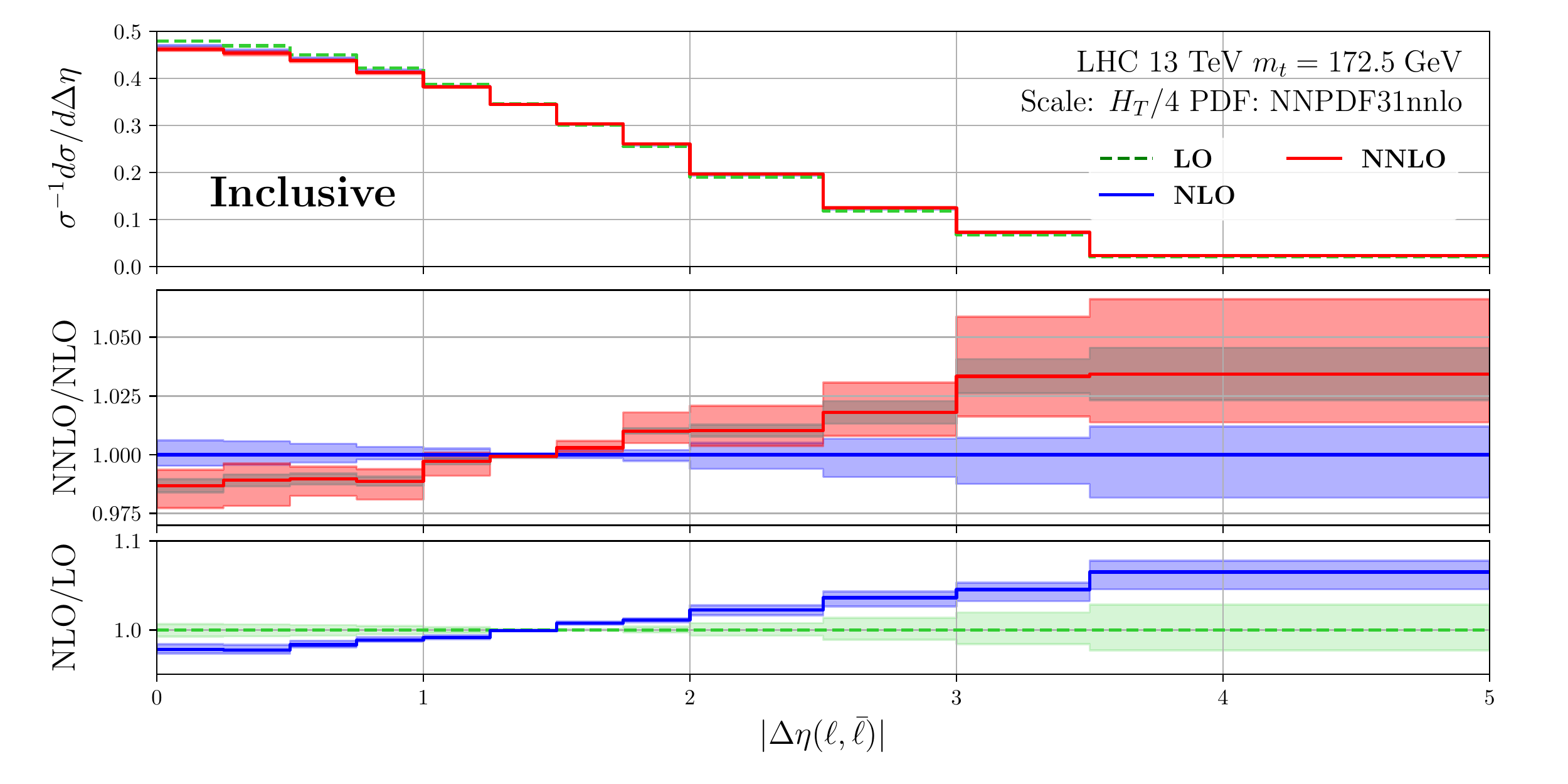}  
   \caption{NNLO QCD predictions for the fiducial (top) and inclusive selections (bottom) of the normalized $\deta$ distribution. Uncertainty bands are from 7-point scale variation.}
   \label{fig:deta}
\end{figure}

The normalized fiducial and inclusive $\dphi$ and $\deta$ distributions are shown in fig.~\ref{fig:dphi} and fig.~\ref{fig:deta}, respectively. Each curve is normalized with respect to the corresponding visible cross-section, i.e. the integral under it equals unity. The $\dphi$ distribution is compared with the published ATLAS data \cite{ATLAS:2018rgl}; the $\deta$ one is not since the corresponding data has not been published yet.

A number of observations can be made from fig.~\ref{fig:dphi}. The most interesting feature is the different behavior of the NNLO/NLO $\dphi$ $K$-factor between the fiducial and inclusive cases. With respect to the inclusive case, in the fiducial case the $K$-factor is much larger, the NNLO distribution is in good agreement with data and the scale uncertainty is much larger. Notably, the NNLO inclusive prediction does not agree well with data.

Since both the fiducial and inclusive data originate from the same measurement it is not a priori clear why the NNLO calculation would agree with only one of them. In our view the most plausible explanation for this discrepancy lies in the extrapolation of the fiducial measurement to the full phase space. 

Such a conclusion should not come as a complete surprise since the extrapolation to full phase space is performed with event generators that have accuracy different than the one in the present work. In fact an early indication about the importance of higher order corrections in top quark production came from the long standing top quark $p_T$ discrepancy, namely, that NLO-accurate event generators do not model well the LHC top quark $p_T$ distribution while the NNLO QCD correction significantly improves the agreement with data. 

\subsection{Anatomy of higher order corrections to $\dphi$}\label{sec:anatomy}

In the following we offer a detailed analysis quantifying a number of possible contributions to this observable. We show that they are too small to affect the behavior of this observable in the SM.

{\it Is the NNLO correction large?} NLO analyses \cite{ATLAS:2018rgl} indicate that higher order effects are likely not going to bridge the $3.2\sigma$ discrepancy with the ATLAS $\dphi$ data. Yet we see that the NNLO QCD prediction agrees well with data in the fiducial region. From this one cannot directly conclude that the NNLO correction is unusually large. The reason is that our NNLO prediction uses scales different than the ones in most event generators. 

For our preferred choice of scales we find that the fiducial NNLO/NLO $K$-factor is no larger than 5\%. This is perfectly reasonable NNLO correction which, moreover, is consistent with the NLO scale uncertainty band. The NLO/LO $K$-factor is larger by a factor of about 3. In the inclusive case one observes smaller $K$-factors and less scale variation which is reasonable to expect since the observable is more inclusive. We note that in both cases the smallness of the LO uncertainty band is due to a cancellation between the normalization factor and is not representative of the true uncertainty in the differential distribution.

We conclude that the behavior of $\dphi$ is consistent with good perturbative convergence.  The NNLO correction plays an important role: in the fiducial case it reduces the scale uncertainty by more than a factor of two and modifies the slope of the theory prediction in a direction that improves the agreement with data.
\begin{figure}[t]
  \centering
     \hspace{-1mm} 
     \includegraphics[width=0.48\textwidth]{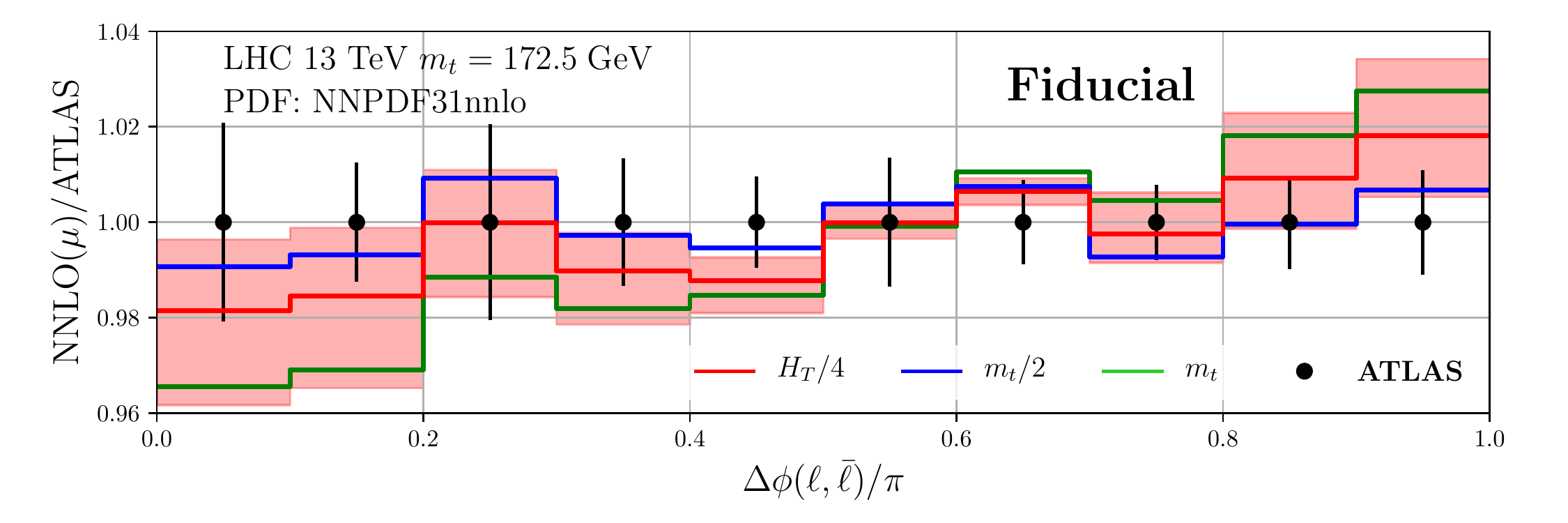}
   \caption{Three NNLO QCD predictions utilizing different scales versus ATLAS data \cite{ATLAS:2018rgl}. The red band represents the 7-point scale variation for the default scale choice eq.~(\ref{eq:scales}).}
   \label{fig:scales}
\end{figure}

{\it Choice of scales.} All calculations in this work are performed with three scales: the one in eq.~(\ref{eq:scales}) as well as $\mu_{F,R}=m_t$ and $\mu_{F,R}=m_t/2$. As can be seen in fig.~\ref{fig:scales} the result with scale $m_t/2$ behaves similarly to the one in eq.~(\ref{eq:scales}) and is even closer to data. On the other hand, the calculation with scale $m_t$ has larger NNLO/NLO $K$-factor and the agreement with data in the fiducial case is not as good as for the other two scales. 

To understand this behavior we recall that the scale $\mu_{F,R}=m_t/2$ was found in ref.~\cite{Czakon:2016dgf} to lead to fast perturbative convergence for the total cross-section. This behavior is similar to the default dynamic scale of eq.~(\ref{eq:scales}). However, perturbative convergence with the canonical scale $\mu_{F,R}=m_t$ is slower. We conclude that the pattern of higher order corrections for the fiducial $\dphi$ distribution is in line with our previous findings for generic top quark differential distributions. We expect that the predictions based on the default dynamic scale as well as on the scale $\mu_{F,R}=m_t/2$ will not have significant corrections beyond NNLO. By contrast, the scale $\mu_{F,R}=m_t$ may lead to non-negligible corrections beyond NNLO which is the reason, we believe, it does not describe data as well.

{\it Value of $m_t$.} With the help of a NLO calculation we have checked that the value of the top quark mass does not affect the $\dphi$ distribution in a significant way. This may be expected on purely dimensional grounds ($\dphi$ is a dimensionless variable). Nevertheless, a dedicated analysis is warranted in light of the findings of ref.~\cite{Frixione:2014ala} where it was found that the treatment of spin correlations in certain lepton distributions does have a substantial impact on the extracted top mass.

{\it PDF dependence.} The effect on the normalized $\dphi$ distribution is at the level of 1\% and thus marginal. We have checked that by comparing two different PDF sets (NNPDF3.1 and CT14), including their PDF errors.

{\it Finite width and Electroweak corrections.} We have performed a qualitative check of these effects at NLO using the results of ref.~\cite{Denner:2016jyo}. While the setup for that reference is different from ours the comparison indicates that the effects on the $\dphi$ distribution are small, perhaps of the order of 1\%. It will be very valuable to investigate such effects in detail in the future.

{\it Top production versus top decay.} We find that radiative corrections to top quark decay have a small impact on the $\dphi$ distribution. 

\subsection{Observables other than inclusive $\dphi$}

Following refs.~\cite{Mahlon:2010gw,ATLAS:2018rgl} we have also investigated the $\dphi$ distribution for several ``slices" of the $t\t$ invariant mass $\Mtt$. Due to space limitation we present no results here (however see \cite{web-based-results}) but only remark that the NNLO corrections are small, in the sense that they are well within the NLO scale uncertainty band and have much reduced scale variation relative to NLO. Interpreting such results is, however, subtle since our definition of $\Mtt$ is based on the true top momenta unlike the experimental setup where the tops are reconstructed (see also refs.~\cite{Mahlon:2010gw,Melnikov:2011ai} for a discussion on this point).

Finally, in fig.~\ref{fig:deta} we show the fiducial and inclusive predictions for the $\deta$ distribution. Unlike the $\dphi$ distribution, the NNLO corrections are significant both in the fiducial and inclusive cases. It will be very interesting to compare these predictions with data once it becomes available. An agreement of our NNLO prediction with future data is likely to validate our interpretation of higher order corrections in the $\dphi$ distribution discussed in sec.~\ref{sec:anatomy}.

\subsection{Quantifying spin correlations}

\begin{figure}[t]
  \centering
     \hspace{-1mm} 
     \includegraphics[width=0.48\textwidth]{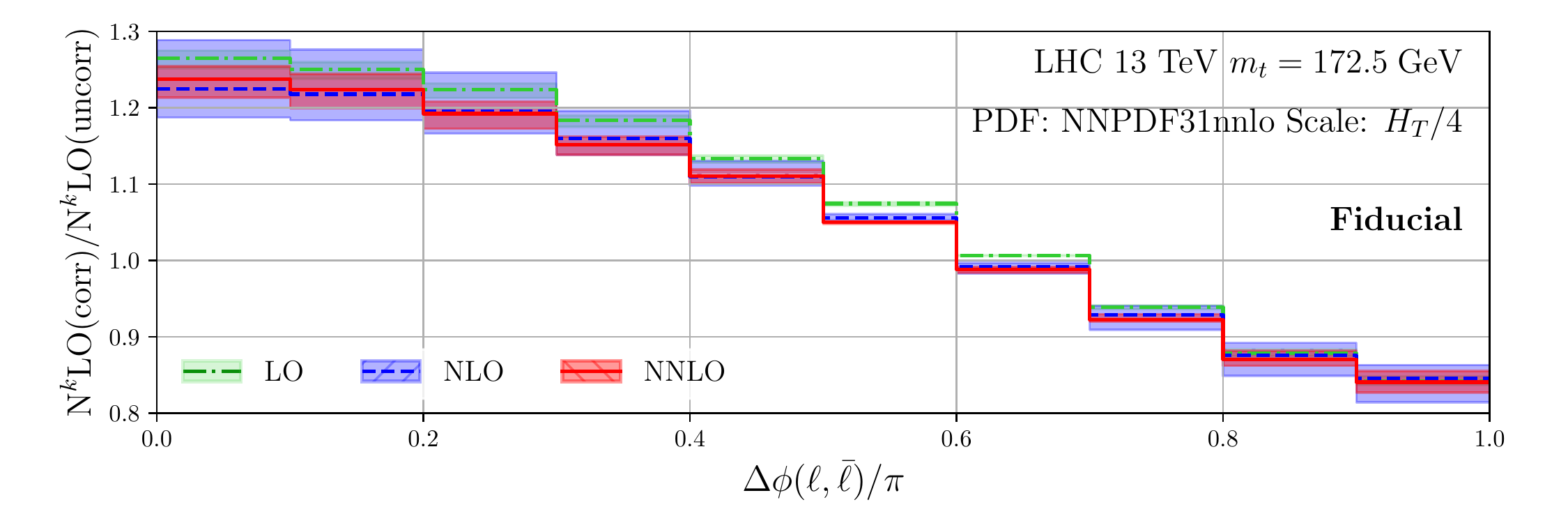}
   \caption{Size of spin correlations in the fiducial $\dphi$ distribution at each order through NNLO in QCD.}
   \label{fig:corr-size}
\end{figure}

In fig.~\ref{fig:corr-size} we show the magnitude of spin correlations in the $\dphi$ distribution through NNLO in QCD. To that end we take the ratio of the calculations with and without spin correlations at a given order. The former calculation is performed by taking spin-averaged top-production times spin-averaged top decay. We observe that spin correlations are large and change little at higher orders.

\begin{figure}[h]
  \centering
     \hspace{-1mm} 
     \includegraphics[width=0.48\textwidth]{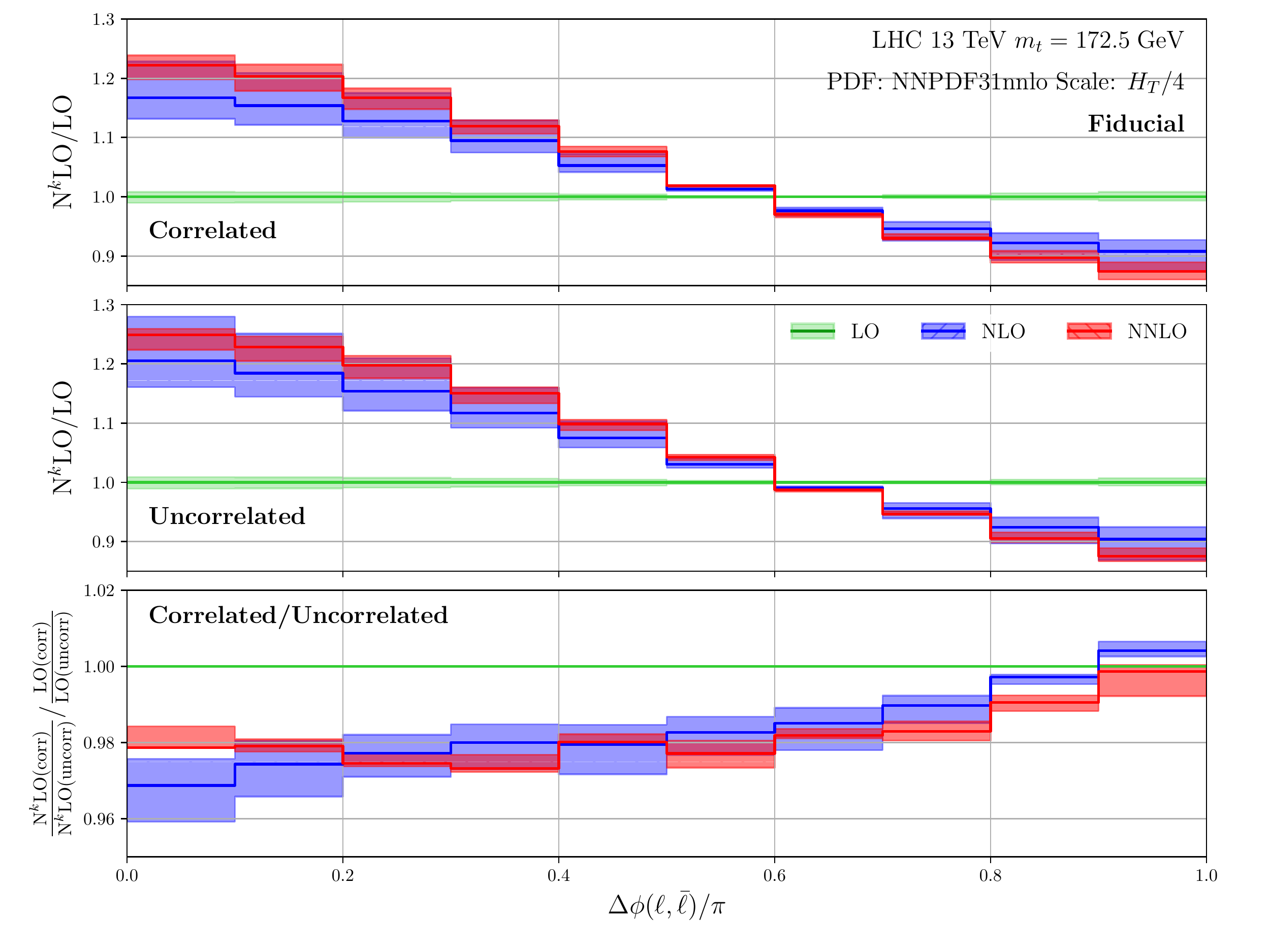}
   \caption{Disentangling radiative corrections from spin correlations for the fiducial $\dphi$ distribution. Shown is the ratio N$^k$LO/LO, for $k=0,1,2$, for the spin-correlated calculation (top), for the calculation without spin correlation (middle) and their ratio (bottom). The bands represent the spread of the ratios for each of the 7 scale variations.}
   \label{fig:corr-dist}
\end{figure}

In order to disentangle the effect of kinematics from spin correlations, in fig.~\ref{fig:corr-dist} we show the ratios NLO/LO and NNLO/LO separately for the exact (top panel) and spin-uncorrelated (middle panel) cases. We observe that all these $K$-factors are significant in size and nearly identical to each other at a given perturbative order. This means that while higher order corrections are substantial they largely decouple from spin correlations. Indeed, the difference between the two NLO/LO and NNLO/LO bands is much smaller than their individual magnitudes. This can be seen more clearly in the bottom panel where their ratio is taken.

Our analysis shows that the control of higher order corrections in the $\dphi$ distribution is essential for interpreting spin correlations with high precision. This is because in this observable spin correlations and kinematics are mixed in a very non-trivial way and therefore a detailed analysis of spin correlations requires good understanding of kinematic effects.

\section{Conclusions}

In this work we compute, for the first time, the complete set of NNLO QCD corrections to top-pair production and decay at hadron colliders. We work in the narrow width approximation for both the top quark and the $W$ boson. We utilize this calculation for the study of spin correlations in top-pair production in the dilepton channel.

Our calculation shows that NNLO QCD corrections to realistic dilepton top quark pair final states play an important role: they increase the SM prediction, significantly decrease the dominant scale uncertainty and improve the agreement with data. 

Using the scales advocated previously in the context of stable top production, we find that NNLO QCD agrees with the recent 13 TeV ATLAS data thus alleviating, or perhaps removing altogether, the earlier reported $3.2\sigma$ discrepancy with respect to the SM. 

An important finding of the present work is that data extrapolation to full phase space with existing event generators seems not to be compatible with the direct NNLO QCD calculation. We believe that thanks to the very high precision of both theory predictions and experimental measurements we begin to see clear evidence that top quark measurements begin to resolve and constrain such delicate modeling effects.

\begin{acknowledgments}
We thank the authors of the {\tt OpenLoops} library for an extensive correspondence and for kindly providing us with a private version of {\tt OpenLoops2}. We also thank M.~Pellen for a cross-check with the results of ref.~\cite{Denner:2016jyo}. A.M. thanks the Department of Physics at Princeton University for hospitality during the completion of this work. The work of A. B. was supported in part by BMBF project 05H18VKCC1. The work of M.C. was supported in part by a grant of the BMBF. The work of A.M., A.P. and R.P. is supported by the European Research Council Consolidator Grant NNLOforLHC2. The work of A.M. and A.P. was also supported by the UK STFC grants ST/L002760/1 and ST/K004883/1.
\end{acknowledgments}


\begin{thebibliography}{99}

\bibitem{Barger:1988jj} 
  V.~D.~Barger, J.~Ohnemus and R.~J.~N.~Phillips,
  Int.\ J.\ Mod.\ Phys.\ A {\bf 4}, 617 (1989).
  
\bibitem{Bigi:1986jk} 
  I.~I.~Y.~Bigi, Y.~L.~Dokshitzer, V.~A.~Khoze, J.~H.~K\"uhn and P.~M.~Zerwas,
  Phys.\ Lett.\ B {\bf 181}, 157 (1986).

\bibitem{Bernreuther:1995cx} 
  W.~Bernreuther, A.~Brandenburg and P.~Uwer,
  Phys.\ Lett.\ B {\bf 368}, 153 (1996)
  [hep-ph/9510300].
  
\bibitem{Mahlon:1995zn} 
  G.~Mahlon and S.~J.~Parke,
  Phys.\ Rev.\ D {\bf 53}, 4886 (1996)
  [hep-ph/9512264].

\bibitem{Bernreuther:2000yn} 
  W.~Bernreuther, A.~Brandenburg and Z.~G.~Si,
  Phys.\ Lett.\ B {\bf 483}, 99 (2000)
  [hep-ph/0004184].

\bibitem{Bernreuther:2001bx} 
  W.~Bernreuther, A.~Brandenburg, Z.~G.~Si and P.~Uwer,
  Phys.\ Lett.\ B {\bf 509}, 53 (2001)
  [hep-ph/0104096].

\bibitem{Bernreuther:2001rq} 
  W.~Bernreuther, A.~Brandenburg, Z.~G.~Si and P.~Uwer,
  Phys.\ Rev.\ Lett.\  {\bf 87}, 242002 (2001)
  [hep-ph/0107086].
      
\bibitem{Mahlon:2010gw} 
  G.~Mahlon and S.~J.~Parke,
  Phys.\ Rev.\ D {\bf 81}, 074024 (2010)
  [arXiv:1001.3422 [hep-ph]].
  
\bibitem{Bernreuther:2010ny} 
  W.~Bernreuther and Z.~G.~Si,
  Nucl.\ Phys.\ B {\bf 837}, 90 (2010)
  [arXiv:1003.3926 [hep-ph]].

\bibitem{Melnikov:2011ai} 
  K.~Melnikov and M.~Schulze,
  Phys.\ Lett.\ B {\bf 700}, 17 (2011)
  [arXiv:1103.2122 [hep-ph]].
    
\bibitem{Bernreuther:2015yna} 
  W.~Bernreuther, D.~Heisler and Z.~G.~Si,
  JHEP {\bf 1512}, 026 (2015)
  [arXiv:1508.05271 [hep-ph]].
    
\bibitem{Bernreuther:2017yhg} 
  W.~Bernreuther, P.~Galler, Z.~G.~Si and P.~Uwer,
  Phys.\ Rev.\ D {\bf 95}, no. 9, 095012 (2017)
  [arXiv:1702.06063 [hep-ph]].

\bibitem{Aguilar-Saavedra:2018ggp} 
  J.~A.~Aguilar-Saavedra,
  JHEP {\bf 1809}, 116 (2018)
  [arXiv:1806.07438 [hep-ph]].

\bibitem{Richardson:2018pvo} 
  P.~Richardson and S.~Webster,
  arXiv:1807.01955 [hep-ph].

\bibitem{Hill:1993hs} 
  C.~T.~Hill and S.~J.~Parke,
  Phys.\ Rev.\ D {\bf 49}, 4454 (1994)
  [hep-ph/9312324].
  
\bibitem{Han:2012fw} 
  Z.~Han, A.~Katz, D.~Krohn and M.~Reece,
  JHEP {\bf 1208}, 083 (2012)
  [arXiv:1205.5808 [hep-ph]].

\bibitem{Baumgart:2012ay} 
  M.~Baumgart and B.~Tweedie,
  JHEP {\bf 1303}, 117 (2013)
  [arXiv:1212.4888 [hep-ph]].
  
\bibitem{Han:2013lna} 
  Z.~Han and A.~Katz,
  arXiv:1310.0356 [hep-ph].

\bibitem{Biswas:2014hwa} 
  S.~Biswas, R.~Frederix, E.~Gabrielli and B.~Mele,
  JHEP {\bf 1407}, 020 (2014)
  [arXiv:1403.1790 [hep-ph]].
  
\bibitem{ATLAS:2018rgl} 
  The ATLAS collaboration [ATLAS Collaboration],
  ATLAS-CONF-2018-027.
  
\bibitem{Chatrchyan:2013wua} 
  S.~Chatrchyan {\it et al.} [CMS Collaboration],
  Phys.\ Rev.\ Lett.\  {\bf 112}, no. 18, 182001 (2014)
  [arXiv:1311.3924 [hep-ex]].
  
\bibitem{Aad:2014pwa} 
  G.~Aad {\it et al.} [ATLAS Collaboration],
  Phys.\ Rev.\ D {\bf 90}, no. 11, 112016 (2014)
  [arXiv:1407.4314 [hep-ex]].

\bibitem{Abazov:2015psg} 
  V.~M.~Abazov {\it et al.} [D0 Collaboration],
  Phys.\ Lett.\ B {\bf 757}, 199 (2016)
  [arXiv:1512.08818 [hep-ex]].
  
\bibitem{Khachatryan:2016xws} 
  V.~Khachatryan {\it et al.} [CMS Collaboration],
  Phys.\ Rev.\ D {\bf 93}, no. 5, 052007 (2016)
  [arXiv:1601.01107 [hep-ex]].
  
\bibitem{Aaboud:2016bit} 
  M.~Aaboud {\it et al.} [ATLAS Collaboration],
  JHEP {\bf 1703}, 113 (2017)
  [arXiv:1612.07004 [hep-ex]].

\bibitem{Czakon:2014xsa} 
  M.~Czakon, P.~Fiedler and A.~Mitov,
  Phys.\ Rev.\ Lett.\  {\bf 115}, no. 5, 052001 (2015)
  [arXiv:1411.3007 [hep-ph]].

\bibitem{Czakon:2015owf} 
  M.~Czakon, D.~Heymes and A.~Mitov,
  Phys.\ Rev.\ Lett.\  {\bf 116}, no. 8, 082003 (2016)
  [arXiv:1511.00549 [hep-ph]].

\bibitem{Gao:2017goi} 
  J.~Gao and A.~S.~Papanastasiou,
  Phys.\ Rev.\ D {\bf 96}, no. 5, 051501 (2017)
  [arXiv:1705.08903 [hep-ph]].

\bibitem{Denner:2012yc} 
  A.~Denner, S.~Dittmaier, S.~Kallweit and S.~Pozzorini,
  JHEP {\bf 1210}, 110 (2012)
  [arXiv:1207.5018 [hep-ph]].
  
\bibitem{Denner:2016jyo} 
  A.~Denner and M.~Pellen,
  JHEP {\bf 1608}, 155 (2016)
  [arXiv:1607.05571 [hep-ph]].

\bibitem{Berger:2016oht} 
  E.~L.~Berger, J.~Gao, C.-P.~Yuan and H.~X.~Zhu,
  Phys.\ Rev.\ D {\bf 94}, no. 7, 071501 (2016)
  [arXiv:1606.08463 [hep-ph]].
  
\bibitem{Berger:2017zof} 
  E.~L.~Berger, J.~Gao and H.~X.~Zhu,
  JHEP {\bf 1711}, 158 (2017)
  [arXiv:1708.09405 [hep-ph]].

\bibitem{Bernreuther:2004jv} 
  W.~Bernreuther, A.~Brandenburg, Z.~G.~Si and P.~Uwer,
  Nucl.\ Phys.\ B {\bf 690}, 81 (2004)
  [hep-ph/0403035].

\bibitem{Melnikov:2009dn} 
  K.~Melnikov and M.~Schulze,
  JHEP {\bf 0908}, 049 (2009)
  [arXiv:0907.3090 [hep-ph]].

\bibitem{Campbell:2012uf} 
  J.~M.~Campbell and R.~K.~Ellis,
  J.\ Phys.\ G {\bf 42}, no. 1, 015005 (2015)
  [arXiv:1204.1513 [hep-ph]].
  
\bibitem{Broggio:2014yca} 
  A.~Broggio, A.~S.~Papanastasiou and A.~Signer,
  JHEP {\bf 1410}, 98 (2014)
  [arXiv:1407.2532 [hep-ph]].

\bibitem{Czakon:2010td} 
  M.~Czakon,
  Phys.\ Lett.\ B {\bf 693}, 259 (2010)
  [arXiv:1005.0274 [hep-ph]].

\bibitem{Czakon:2011ve} 
  M.~Czakon,
  Nucl.\ Phys.\ B {\bf 849}, 250 (2011)
  [arXiv:1101.0642 [hep-ph]].
  
\bibitem{Czakon:2014oma} 
  M.~Czakon and D.~Heymes,
  Nucl.\ Phys.\ B {\bf 890}, 152 (2014)
  [arXiv:1408.2500 [hep-ph]].

\bibitem{Catani:1996jh} 
  S.~Catani and M.~H.~Seymour,
  Phys.\ Lett.\ B {\bf 378}, 287 (1996)
  [hep-ph/9602277].

\bibitem{Catani:2002hc} 
  S.~Catani, S.~Dittmaier, M.~H.~Seymour and Z.~Trocsanyi,
  Nucl.\ Phys.\ B {\bf 627}, 189 (2002)
  [hep-ph/0201036].
  
\bibitem{Campbell:2004ch} 
  J.~M.~Campbell, R.~K.~Ellis and F.~Tramontano,
  Phys.\ Rev.\ D {\bf 70}, 094012 (2004)
  [hep-ph/0408158].
  
\bibitem{Czakon:2016ckf} 
  M.~Czakon, P.~Fiedler, D.~Heymes and A.~Mitov,
  JHEP {\bf 1605}, 034 (2016)
  [arXiv:1601.05375 [hep-ph]].

\bibitem{Czakon:2016dgf} 
  M.~Czakon, D.~Heymes and A.~Mitov,
  JHEP {\bf 1704}, 071 (2017)
  [arXiv:1606.03350 [hep-ph]].

\bibitem{upcomingOL2} 
  F.~Buccioni, J.~N.~Lang, J.~M.~Lindert, P.~Maierh\"ofer, S.~Pozzorini, H.~Zhang, M.~Zoller, 
  {\tt OpenLoops2}, to appear.
  
\bibitem{Cascioli:2011va} 
  F.~Cascioli, P.~Maierh\"ofer and S.~Pozzorini,
  Phys.\ Rev.\ Lett.\  {\bf 108}, 111601 (2012)
  [arXiv:1111.5206 [hep-ph]].

\bibitem{Buccioni:2017yxi} 
  F.~Buccioni, S.~Pozzorini and M.~Zoller,
  Eur.\ Phys.\ J.\ C {\bf 78}, no. 1, 70 (2018)
  [arXiv:1710.11452 [hep-ph]].
  
\bibitem{Chen:2017jvi} 
  L.~Chen, M.~Czakon and R.~Poncelet,
  JHEP {\bf 1803}, 085 (2018)
  [arXiv:1712.08075 [hep-ph]].

\bibitem{Baernreuther:2013caa} 
  P.~B\"arnreuther, M.~Czakon and P.~Fiedler,
  JHEP {\bf 1402}, 078 (2014)
  [arXiv:1312.6279 [hep-ph]].

\bibitem{Bonciani:2008wf} 
  R.~Bonciani and A.~Ferroglia,
  JHEP {\bf 0811}, 065 (2008)
  [arXiv:0809.4687 [hep-ph]].

\bibitem{Asatrian:2008uk} 
  H.~M.~Asatrian, C.~Greub and B.~D.~Pecjak,
  Phys.\ Rev.\ D {\bf 78}, 114028 (2008)
  [arXiv:0810.0987 [hep-ph]].
  
\bibitem{Beneke:2008ei} 
  M.~Beneke, T.~Huber and X.-Q.~Li,
  Nucl.\ Phys.\ B {\bf 811}, 77 (2009)
  [arXiv:0810.1230 [hep-ph]].

\bibitem{Lim:2018qiw} 
  M.~A.~Lim,
  arXiv:1811.10874 [hep-ph].

\bibitem{Cullen:2014yla} 
  G.~Cullen {\it et al.},
  Eur.\ Phys.\ J.\ C {\bf 74}, no. 8, 3001 (2014)
  [arXiv:1404.7096 [hep-ph]].

\bibitem{Gao:2012ja} 
  J.~Gao, C.~S.~Li and H.~X.~Zhu,
  Phys.\ Rev.\ Lett.\  {\bf 110}, no. 4, 042001 (2013)
  [arXiv:1210.2808 [hep-ph]].
  
\bibitem{Brucherseifer:2013iv} 
  M.~Brucherseifer, F.~Caola and K.~Melnikov,
  JHEP {\bf 1304}, 059 (2013)
  [arXiv:1301.7133 [hep-ph]].

\bibitem{Blokland:2004ye} 
  I.~R.~Blokland, A.~Czarnecki, M.~Slusarczyk and F.~Tkachov,
  Phys.\ Rev.\ Lett.\  {\bf 93}, 062001 (2004)
  [hep-ph/0403221].
  
\bibitem{Ball:2017nwa} 
  R.~D.~Ball {\it et al.} [NNPDF Collaboration],
  Eur.\ Phys.\ J.\ C {\bf 77}, no. 10, 663 (2017)
  [arXiv:1706.00428 [hep-ph]].
  
\bibitem{Dulat:2015mca} 
  S.~Dulat {\it et al.},
  Phys.\ Rev.\ D {\bf 93}, no. 3, 033006 (2016)
  [arXiv:1506.07443 [hep-ph]].

\bibitem{Kluge:2006xs} 
  T.~Kluge, K.~Rabbertz and M.~Wobisch,
  hep-ph/0609285.

\bibitem{Britzger:2012bs} 
  D.~Britzger {\it et al.} [fastNLO Collaboration],
  arXiv:1208.3641 [hep-ph].

\bibitem{Czakon:2017dip} 
  M.~Czakon, D.~Heymes and A.~Mitov,
  arXiv:1704.08551 [hep-ph].

\bibitem{web-based-results}
 Repository with results and additional plots: 
 \url{http://www.precision.hep.phy.cam.ac.uk/results/ttbar-decay/}

\bibitem{Cacciari:2008gp} 
  M.~Cacciari, G.~P.~Salam and G.~Soyez,
  JHEP {\bf 0804}, 063 (2008)
  [arXiv:0802.1189 [hep-ph]].

\bibitem{Frixione:2014ala} 
  S.~Frixione and A.~Mitov,
  JHEP {\bf 1409}, 012 (2014)
  [arXiv:1407.2763 [hep-ph]].
                     
\end{thebibliography}
\end{document}